\documentclass[10pt,aps,prl,twocolumn,superscriptaddress,showpacs,preprintnumbers,amsmath,amssymb]{revtex4-1}

\usepackage{amsmath,amssymb,amsfonts}
\usepackage{graphicx,color}
\usepackage{dcolumn}
\usepackage{bm}
\usepackage{natbib}
\usepackage{verbatim}
\usepackage[ruled,lined,titlenotnumbered]{algorithm2e}

\begin{document}

\title{Network reconstruction via density sampling}

\author{Tiziano Squartini}
\email{tiziano.squartini@imtlucca.it}
\affiliation{IMT School for Advanced Studies, Piazza S.Francesco 19, 55100 Lucca - Italy}
\author{Giulio Cimini}
\affiliation{IMT School for Advanced Studies, Piazza S.Francesco 19, 55100 Lucca - Italy}
\affiliation{Istituto dei Sistemi Complessi (ISC) - CNR, UoS Sapienza, Dipartimento di Fisica, Universit\'a ``Sapienza'', Piazzale Aldo Moro 5, 00185 Rome - Italy}
\author{Andrea Gabrielli}
\affiliation{IMT School for Advanced Studies, Piazza S.Francesco 19, 55100 Lucca - Italy}
\affiliation{Istituto dei Sistemi Complessi (ISC) - CNR, UoS Sapienza, Dipartimento di Fisica, Universit\'a ``Sapienza'', Piazzale Aldo Moro 5, 00185 Rome - Italy}
\author{Diego Garlaschelli}
\affiliation{Instituut-Lorentz for Theoretical Physics, Leiden Institute of Physics, University of Leiden, Niels Bohrweg 2, 2333 CA Leiden - The Netherlands}
\date{\today}

\begin{abstract}
Reconstructing weighted networks from partial information is necessary in many important circumstances, e.g. for a correct estimation of systemic risk. It has been shown that, in order to achieve an accurate reconstruction, it is crucial to reliably replicate the empirical degree sequence, which is however unknown in many realistic situations. More recently, it has been found that the knowledge of the degree sequence can be replaced by the knowledge of the strength sequence, which is typically accessible, complemented by that of the total number of links, thus considerably relaxing the observational requirements. Here we further relax these requirements and devise a procedure valid when even the the total number of links is unavailable. We assume that, apart from the heterogeneity induced by the degree sequence itself, the network is homogeneous, so that its (global) link density can be estimated by sampling subsets of nodes with representative density. We show that the best way of sampling nodes is the random selection scheme, any other procedure being biased towards unrealistically large, or small, link densities. We then introduce our core technique for reconstructing both the topology and the link weights of the unknown network in detail. When tested on real economic and financial data sets, our method achieves a remarkable accuracy and is very robust with respect to the sampled subsets, thus representing a reliable practical tool whenever the available topological information is restricted to small portions of nodes.
\end{abstract}

\maketitle 

\section*{Introduction}

Reconstructing a weighted, directed network means providing an algorithm to estimate the presence and the weight of all links in the network, making optimal use of the available information \cite{Wells2004,Upper2011,Mastromatteo2012,Baral2012,Drehmann2013,Halaj2013,Anand2014,Montagna2014,Peltonen2015,Cimini2015a}. Since several networks are in general compatible with the known information, the output of such a procedure cannot identify a unique network but rather an ensemble of possible ones. This leads to a (large) set of candidate networks to be sampled with a certain probability, where the latter has to be specified in such a way that the resulting ensemble average is as close as possible to the empirical, unknown network. Maximum-entropy is a powerful method to construct probability distributions that realise a certain set of constraints on average. Treating the available pieces of information as empirical constraints in the maximum-entropy procedure ensures that the statistical inference carried out via the resulting distribution is maximally unbiased.

In many situations, e.g. for economic, interbank or other financial networks, the strength sequence (i.e. the list of strengths of all nodes) is known while there is little or no information available about the topology (i.e. the binary structure) of the network. Exploiting the strength sequence as the only constraint of the maximum entropy procedure leads to an unrealistic ensemble where the likely networks are (almost) completely connected \cite{Mastrandrea2014}. This occurs because, when replicating the empirical strengths in absence of topological information, the method tends to distribute non-zero link weights as evenly as possible (i.e. between all pairs of nodes). When such unrealistically dense networks are used as proxies to measure, e.g. the level of systemic risk in a financial network, the resulting estimates are completely unreliable. By contrast, it has been shown that, if the degree sequence is known in addition to the strength sequence, the network reconstruction procedure improves tremendously and achieves a remarkable accuracy, as a result of a much more faithful replication of the topology \cite{Mastrandrea2014,Cimini2015b}. Notice that, if the link weights are specified by the matrix $\mathbf{W}$, whose entry $w_{ij}\geq 0$ represents the weight of the directed link from node $i$ to node $j$, the topology is specified by the binary adjacency matrix $\mathbf{A}$ whose entry $a_{ij}=1$ if $w_{ij}$ is strictly positive and zero otherwise.

Although complete information on the degree sequence is rarely available, this kind of information can be retrieved from the strength sequence, provided that the latter is complemented with some kind of topological information: in \cite{Musmeci2013} this information consists of the degree sequence of only a subset $I$ of nodes, $\{k_i\}_{i\in I}$, while in \cite{Cimini2015a} the information used is the total number of links, $L$, of the network.

In this paper we face the problem of reconstructing weighted, directed networks, for which the only information available is represented by the set of out-strengths $s^{out}_i=\sum_{j(\neq i)}w_{ij}$ and in-strengths $s^{in}_i=\sum_{j(\neq i)}w_{ji}$ (i.e. the total rows and columns sums of the adjacency matrix) as well as the link density of a subset $I$ of nodes, i.e. $c_I=\frac{L_I}{n_I(n_I-1)}$, with $L_I=\sum_{i\in I}\sum_{j(\neq i)\in I}a_{ij}$ being the observed number of internal links to the subset $I$. By doing so, we do not require information which is either too detailed (as the degree sequence of even a small subset of nodes) or simply unaccessible (as the total number of links). However, the information encoded into the link density of the chosen subset must be representative of the global one, in order to accurately reconstruct a given network: for this reason, we also propose a recipe about how properly sampling the nodes set of our network. As we will show, the random-nodes sampling scheme provides the best way to draw representative subsets out of the whole nodes set.

Concerning the reconstruction of the weighted structure, we will employ the degree-corrected gravity model \cite{Cimini2015a} with a correction term ensuring that the strengths are reproduced even in absence of self-loops, i.e. of diagonal terms indicating self-interactions. As we will show, such a correction becomes more and more important as the strength of the considered node is increased, whence the need to properly account for it.

The rest of the paper is organized as follows. In section ``Methods'' we illustrate the two steps characterizing our reconstruction method and provide measures to test the effectiveness of the algorithm; in section ``Results'' we apply our method to two real networks, an economic one and a financial one, and in section ``Conclusions'' we discuss the results.

\section*{Methods}

\subsection*{Inferring the topological structure}

In order to reconstruct the topological structure of a network $\mathbf{W}$, whenever the nodes strengths $\{s^{out}_i\}_{i=1}^N$ and $\{s^{in}_i\}_{i=1}^N$ and the total number of links $L$ are known, 
one can follow the algorithm proposed in \cite{Cimini2015a}, which prescribes to solve the equation

\begin{equation}\label{eq1}
L=\langle L\rangle
\end{equation}
with $L=\sum_{i}\sum_{j(\neq i)}a_{ij}$, $\langle L\rangle=\sum_{i}\sum_{j(\neq i)}p_{ij}$ and $p_{ij}=(zs^{out}_is^{in}_j)/(1+zs^{out}_is^{in}_j)$, in order to estimate the unknown parameter $z$ and quantify the probability $p_{ij}$ that a directed link from $i$ to $j$ exists. However, a global (yet very simple) piece of information as $L$ may be not always available. In these cases, an algorithm resorting upon local information has to be employed. In this paper we propose an algorithm to infer the unknown parameter $z$ whenever the information of only a subset $I$ of nodes is accessible. Notice that a possible solution to this problem has already been provided in \cite{Musmeci2013}, where the supposedly known piece of information is represented by the degree sequence of the nodes in $I$, i.e. $\{k_i\}_{i\in I}$, an hypothesis leading to the equation

\begin{equation}
\sum_{i\in I}\left(k^{out}_i+k^{in}_i\right)=\sum_{i\in I}\left(\langle k^{out}_i\rangle+\langle k^{in}_i\rangle\right)
\end{equation}
(with $\langle k^{out}_i\rangle=\sum_{j(\neq i)\in V}p_{ij}$ and $\langle k^{in}_i\rangle=\sum_{j(\neq i)\in V}p_{ji}$ and $V$ indicating the whole nodes set). However, the knowledge of the number of neighbors of even a small subset of nodes may be unavailable as well. For this reason, here we make use of a simpler, more easily accessible, information and suppose to know only the link density within the subset $I$. Our recipe thus reads

\begin{equation}\label{eq3}
c_I=\langle c_I\rangle
\end{equation}
where $c_I=L_I/[n_I(n_I-1)]$, $n_I=|I|$ is the number of nodes constituting the subset $I$, $L_I=\sum_{i\in I}\sum_{j(\neq i)\in I}a_{ij}$ is the observed number of links within it and $\langle L_I\rangle=\sum_{i\in I}\sum_{j(\neq i)\in I}p_{ij}$ is the expected value of $L_I$.

Remarkably, eq.(\ref{eq3}) can be easily extended to infer the structure of a different subset (say $I'$), provided that the link density of the latter could be guessed from the known value $c_I$. As an example, let us assume the existence of a linear proportionality between the two values $c_{I'}$ and $c_I$: in this case, the equation to be solved would be

\begin{equation}\label{eq4}
c_I=f\langle c_{I'}\rangle.
\end{equation}

More explicitly, such a condition translates into the equation

\begin{equation}\label{eq5}
c_I=\frac{f}{n_{I'}(n_{I'}-1)}\sum_{i\in I'}\sum_{j(\neq i)\in I'}\frac{z_{I'}s^{out}_is^{in}_j}{1+z_{I'}s^{out}_is^{in}_j}
\end{equation}
which shows that the observed quantity tuning the parameter $z_{I'}$ is $c_I\cdot n_{I'}(n_{I'}-1)$, i.e. the link density of the known subset, corrected by a volume term.

The value $f=1$ corresponds to the assumption that the network is homogeneous. This is equivalent to requiring that any two different subsets have exactly the same link density and that, in turn, any subset provides a representative value of the global network density. As we will show in what follows, a random sampling of the set of nodes indeed ensures that this assumption is verified with high accuracy, for the networks considered here.

\subsection*{Inferring the weighted structure}

Beside reconstructing a network topological features, the approach proposed in \cite{Cimini2015a} satisfactorily reproduces also its weighted structure. This approach is based on the degree-corrected gravity model prescription, which reads

\begin{equation}
w_{ij} = \left\{ \begin{array}{cl}
0 & \textrm{with probability $1-p_{ij}$},\\
\frac{s_i^{out}s_j^{in}}{Wp_{ij}} & \textrm{with probability $p_{ij}$}
\end{array} \right.
\end{equation}
leading to the expectations $\langle w_{ij}\rangle=s^{out}_is^{in}_j/W$ and ensuring that $s^{out}_i=\langle s^{out}_i\rangle=\sum_{j}w_{ij},\:\forall\:i$ and $s^{in}_i=\langle s^{in}_i\rangle=\sum_{j}w_{ji},\:\forall\:i$ (i.e. that the in-strength and out-strength sequences are, on average, reproduced) as long as {\it all} entries are summed over.

However, in many real-world networks self-loops are either absent or explicitly excluded: this implies that either the diagonal terms of the adjacency matrix are equal to zero or that our sums should run over $j\neq i$. This causes the expectations coming from the degree-corrected gravity model to need an extra-term to restore the correct value. More explicitly,

\begin{equation}
\langle s_i^{out}\rangle=\sum_{j(\neq i)}\langle w_{ij}\rangle=\frac{s_i^{out}(W-s_i^{in})}{W}=s_i^{out}-\frac{s_i^{out}s_i^{in}}{W},
\end{equation}
\begin{equation}
\langle s_i^{in}\rangle=\sum_{j(\neq i)}\langle w_{ji}\rangle=\frac{s_i^{in}(W-s_i^{out})}{W}=s_i^{in}-\frac{s_i^{out}s_i^{in}}{W}
\end{equation}
and the missing term to be added up to our expectations is precisely the diagonal term, i.e. $\langle w_{ii}\rangle$. 

Here we provide a solution to the problem above, by redistributing the diagonal term $\langle w_{ii}\rangle$ across the $N-1$ entries of the $i$th row and the $N-1$ entries of the $i$th column. In order to implement it, a procedure inspired to the iterative proportional fitting (IPF) algorithm \cite{bishop2007discrete} can be devised. More specifically, redistributing the diagonal terms across the corresponding rows and columns amounts to redistribute the strengths of the following matrix on the entries equal to 1. Notice that we need to explicitly distinguish the strengths along rows and columns, since the generic weight $w_{ij}$ needs a correction affecting both $i$ and $j$.

\begin{equation}
\begin{minipage}{0.5\textwidth}
	\[
	\begin{array}{ccccc|c}
	0 & 1 & 1 & 1 & \hdots & \frac{s^{out}_1s^{in}_1}{W} \\
	1 & 0 & 1 & 1 & \hdots & \frac{s^{out}_2s^{in}_2}{W} \\
            1 & 1 & 0 & 1 & \hdots & \frac{s^{out}_3s^{in}_3}{W} \\
	1 & 1 & 1 & 0 & \hdots & \frac{s^{out}_4s^{in}_4}{W} \\
	\vdots & \vdots & \vdots & \vdots & \ddots & \vdots \\
	\hline
	\frac{s^{out}_1s^{in}_1}{W} & \frac{s^{out}_2s^{in}_2}{W} & \frac{s^{out}_3s^{in}_3}{W} & \frac{s^{out}_4s^{in}_4}{W} & \hdots & \\
	\end{array} 
	\]
\end{minipage}
\end{equation}

\begin{table*}[t!]
\[\begin{array}{l|c|c|c|c|c}
\hline
\hline
\text{\bf WTW} & n=5\:(\text{CI}\:95\%) & n=10\:(\text{CI}\:95\%) & n=20\:(\text{CI}\:95\%) & n=50\:(\text{CI}\:95\%) & n=100\:(\text{CI}\:95\%) \\
\hline
\hline
\text{2000 - True positive rate} & 0.794\:[0.772,0.816] & 0.779\:[0.765,0.793] & 0.804\:[0.796,0.812] & 0.801\:[0.797,0.806] & 0.801\:[0.799,0.804] \\
\hline
\text{2000 - Specificity} & 0.700\:[0.669,0.731] & 0.742\:[0.726,0.758] & 0.721\:[0.710,0.731] & 0.728\:[0.723,0.734] & 0.729\:[0.726,0.733] \\
\hline
\text{2000 - Positive predicted value} & 0.796\:[0.784,0.808] & 0.810\:[0.803,0.817] & 0.799\:[0.795,0.804] & 0.802\:[0.800,0.805] & 0.802\:[0.801,0.803] \\
\hline
\text{2000 - Accuracy} & 0.755\:[0.750,0.760] & 0.763\:[0.762,0.766] & 0.769\:[0.768,0.770] & 0.771 & 0.771 \\
\hline
\text{2000 - Cosine similarity} & 0.712 & 0.712 & 0.712 & 0.712 & 0.712 \\
\hline
\hline
\text{\bf e-MID} & n=5\:(\text{CI}\:95\%) & n=10\:(\text{CI}\:95\%) & n=20\:(\text{CI}\:95\%) & n=50\:(\text{CI}\:95\%) & n=100\:(\text{CI}\:95\%) \\
\hline
\hline
\text{1999 - True positive rate} & 0.641\:[0.601,0.673] & 0.633\:[0.614,0.653] &  0.633\:[0.620,0.646] &  0.637\:[0.623,0.643] & 0.636\:[0.632,0.640] \\
\hline
\text{1999 - Specificity} & 0.839\:[0.823,0.856] &  0.856\:[0.848,0.864] &  0.860\:[0.854,0.865] & 0.860\:[0.857,0.863] & 0.861\:[0.859,0.862] \\
\hline
\text{1999 - Positive predicted value} & 0.623\:[0.611,0.637] &  0.632\:[0.625,0.639] &  0.633\:[0.628,0.638] & 0.632\:[0.623,0.635] & 0.633\:[0.631,0.634] \\
\hline
\text{1999 - Accuracy} & 0.785\:[0.780,0.790] &  0.795\:[0.794,0.796] &  0.798\:[0.797,0.799] & 0.799\:[0.798,0.800] & 0.799 \\
\hline
\text{1999 - Cosine similarity} & 0.810\:[0.805,0.815] &  0.814\:[0.811,0.816] &  0.816\:[0.815,0.817] & 0.817 & 0.817 \\
\hline
\end{array}\]
\caption{Statistical indicators used to evaluate the performance of our sampled-based reconstruction method, for different cardinalities $n$ of the known subset $I$. Results are shown together with the $95\%$ confidence intervals (not shown whenever their difference affects the significant digits beyond the third one). The considered cardinalities $n=5, 10, 20, 50, 100$ correspond to percentages ranging from $\simeq2\%$ to $\simeq50\%$ of the total number of nodes. As reference values, the link density is $c=0.578$ for the WTW (in the year 2000) and $c=0.274$ for e-MID (in the year 1999).}\label{tab2}
\end{table*}

In order to achieve the aforementioned redistribution, one can compute the iterations of the IPF algorithm

\begin{equation}
\left\{ \begin{array}{cl}
w_{ij}^{(n)}&=\frac{s^{out}_is^{in}_i}{W}\left(\frac{w_{ij}^{(n-1)}}{\sum_{k(\neq i)}w_{ik}^{(n-1)}}\right)\\
w_{ij}^{(n+1)}&=\frac{s^{out}_js^{in}_j}{W}\left(\frac{w_{ij}^{(n)}}{\sum_{k(\neq j)}w_{kj}^{(n)}}\right)
\end{array} \right.
\end{equation}
upon setting the matrix defined by $w_{ij}^{(0)}=1,\:\forall\:i\neq j$ as the initial configuration. As a consequence, we need to correct our probabilistic recipe as

\begin{equation}\label{eq11}
w_{ij} = \left\{ \begin{array}{cl}
0 & \textrm{with probability $1-p_{ij}$},\\
\left(\frac{s_i^{out}s_j^{in}}{W}+w_{ij}^{(\infty)}\right)\frac{1}{p_{ij}} & \textrm{with probability $p_{ij}$}.
\end{array} \right.
\end{equation}

For all practical purposes, a small number of iterations is often enough to achieve a satisfactory degree of accuracy. Here we explicitly report the analytical functional form of the first three IPF algorithm iterations only:

\begin{eqnarray}
w_{ij}^{(1)}&=&\frac{s^{out}_is^{in}_i}{W}\left[\frac{1}{N-1}\right];\nonumber\\
w_{ij}^{(2)}&=&\frac{s^{out}_is^{in}_i}{W}\left[\frac{s^{out}_js^{in}_j}{\sum_{l(\neq j)}s^{out}_ls^{in}_l}\right];\\
w_{ij}^{(3)}&=&\frac{s^{out}_is^{in}_i}{W}\left[\frac{s^{out}_js^{in}_j}{\sum_{l(\neq j)}s^{out}_ls^{in}_l}\right]\left[\frac{1}{\sum_{k(\neq i)}\frac{s^{out}_ks^{in}_k}{\sum_{m(\neq k)}s^{out}_ms^{in}_m}}\right].\nonumber
\end{eqnarray}

A pseudo-code summarizing the two main steps of our algorithm (i.e. eq.(\ref{eq3}) and eq.(\ref{eq11})) is provided in Appendix.

\subsection*{Testing our reconstruction algorithm}

An algorithm aiming at reconstructing the topological structure of a network is an example of a binary classificator which tries to infer whether each link is present or not. In order to test the performance of our reconstruction method we, thus, consider four indicators: the number of {\it true positives}, {\it true negatives}, {\it false positives} and {\it false negatives}. In network terms, the expectation value of such indices reads $\langle TP\rangle=\sum_{i}\sum_{j(\neq i)}a_{ij}p_{ij}$, $\langle TN\rangle=\sum_{i}\sum_{j(\neq i)}(1-a_{ij})(1-p_{ij})$, $\langle FP\rangle=\sum_{i}\sum_{j(\neq i)}(1-a_{ij})p_{ij}$ and $\langle FN\rangle=\sum_{i}\sum_{j(\neq i)}a_{ij}(1-p_{ij})$. However, the information provided by these indicators is often condensed into four alternative indices. The first one is called {\it sensitivity} (or {\it true positive rate}), $\langle TPR\rangle=\frac{\langle TP\rangle}{L}$, and quantifies the percentage of 1s that are correctly recovered by our method. The second index is the {\it specificity} (or {\it true negative rate}), $\langle SPC\rangle=\frac{\langle TN\rangle}{N(N-1)-L}$, and quantifies the percentage of 0s that are correctly recovered by our method. The third index is the {\it precision} (or {\it positive predicted value}), $\langle PPV\rangle=\frac{\langle TP\rangle}{\langle L\rangle}$, and measures the performance of our method in correctly placing the 1s with respect to the total number of predicted 1s. The fourth index is the {\it accuracy}, $\langle ACC\rangle=\frac{\langle TP\rangle+\langle TN\rangle}{N(N-1)}$, and quantifies the overall performance of our method in correctly placing both the 1s and the 0s.

To test the effectiveness of the weighted reconstruction, instead, we use the cosine similarity measure which estimates the distance between the observed weights $\{w_{ij}\}_{i,j=1}^N$ and the conditional expected weights under our model $\{\langle w_{ij}|a_{ij}=1\rangle\}_{i,j=1}^N$ by treating the corresponding matrices as vectors of real numbers and measuring their overlap. In formulas,

\begin{equation}
\theta=\frac{\mathbf{W}\cdot \langle\mathbf{W}\rangle}{||\mathbf{W}||\:||\langle\mathbf{W}\rangle||}
\end{equation}
with $\theta=-1$ indicating maximum dissimilarity, $\theta=0$ indicating absence of correlations and $\theta=1$ indicating perfect overlap.

\begin{figure*}[t!]
\includegraphics[width=0.8\textwidth]{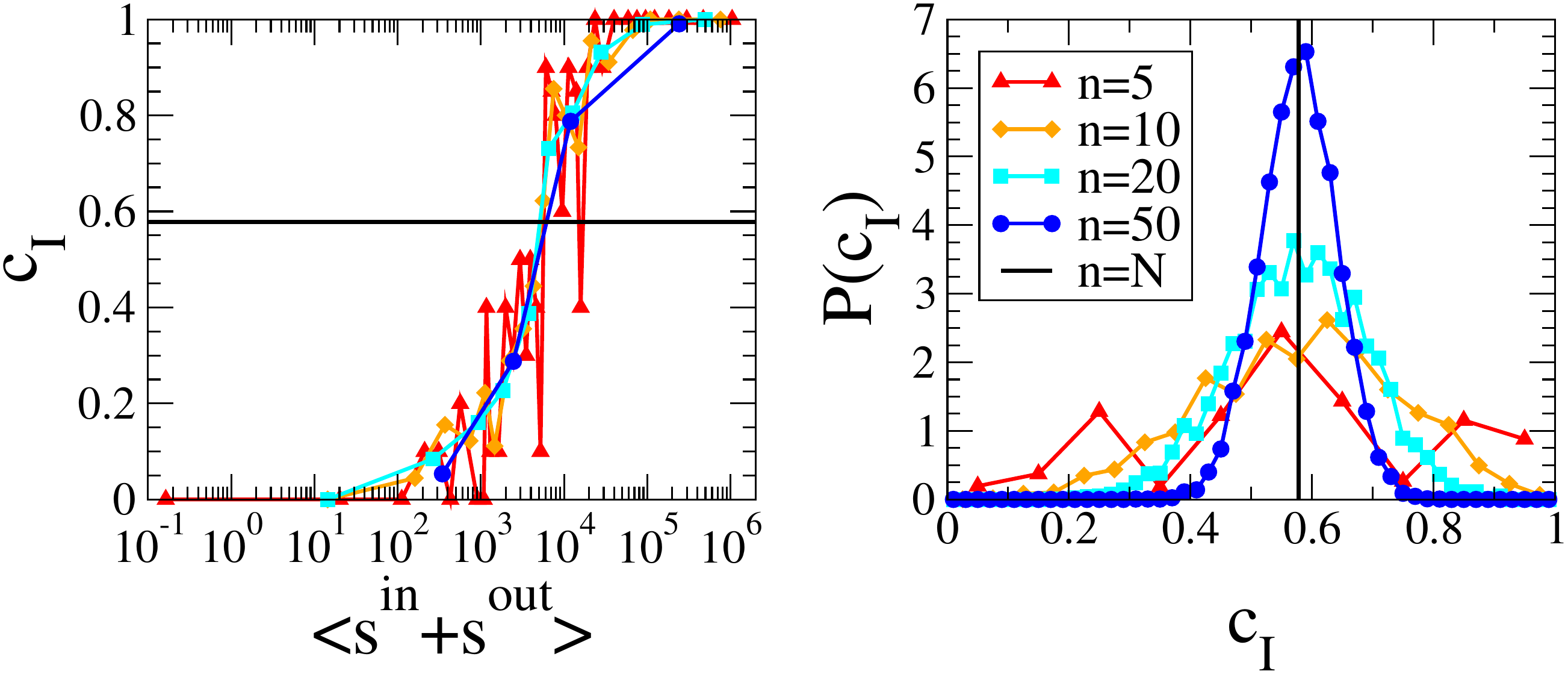}
\includegraphics[width=0.8\textwidth]{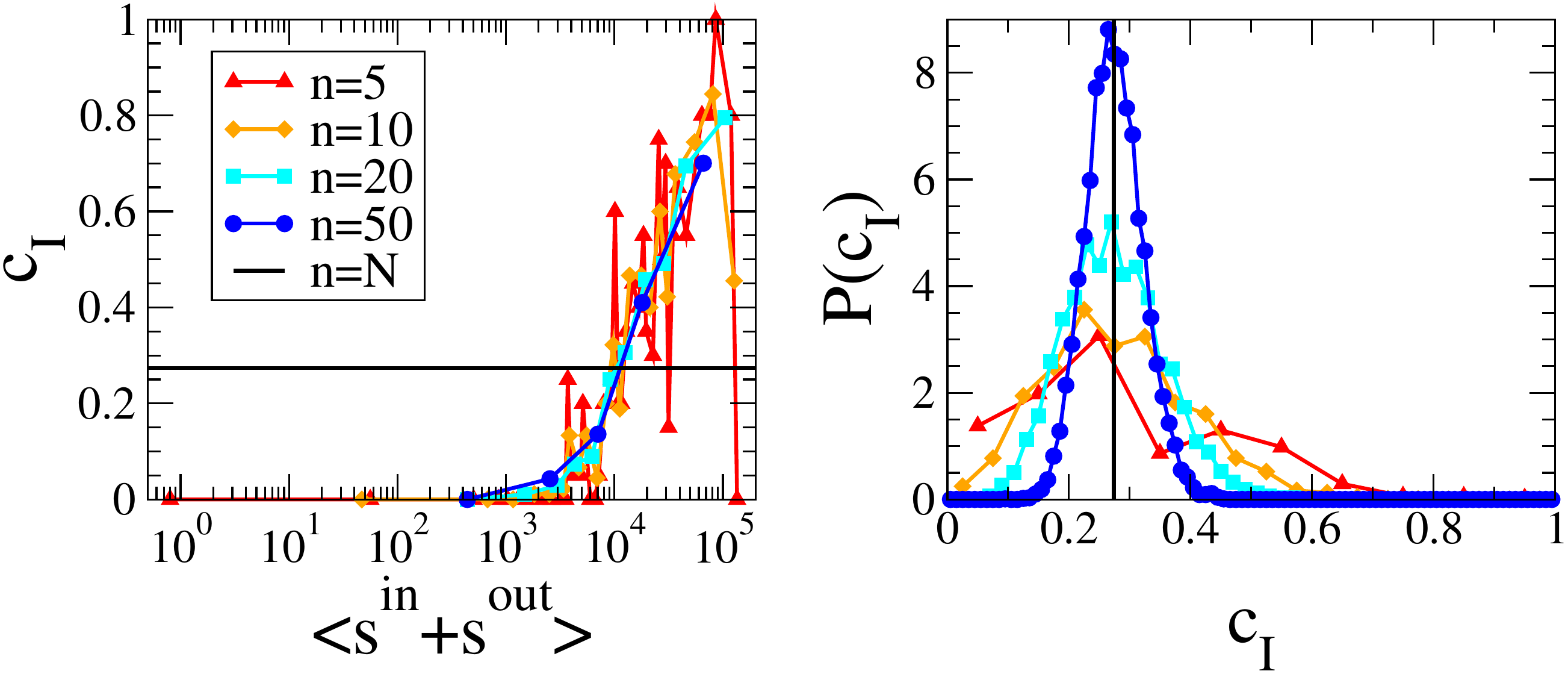}
\caption{Left panels: scatter plots of the link density $c_I$ versus the internal total strength $s^{tot}_I$ of the subset $I$. Nodes characterized by large values of the total strength tend to form densely-connected groups, while nodes characterized by small values of the total strength tend, on the contrary, to form loosely-connected groups. Right panels: empirical probability distributions of the link density $c_I$, when nodes belonging to $I$ are chosen randomly. Each distribution is peaked around the density value of the whole network. Top panels refer to the WTW, bottom panels to e-MID. \label{fig1}}
\end{figure*}

\section*{Results}

\subsection*{World Trade Web}

The first network we have analyzed is the World Trade Web (WTW), i.e. the network whose nodes are the world countries and whose links represent the trade volumes between them: in other words, $w_{ij}$ quantifies the volume of export from $i$ to $j$. We remand the reader to \cite{gleditsch2002expanded} for more details on the dataset. For the sake of illustration, we show detailed results for the snapshot of the WTW in year 2000. We have however analyzed other temporal snapshots as well and found comparable results (see Appendix).

Table \ref{tab2} sums up the results of our analysis when the nodes subset $I$ is chosen at random. We see that the performance of our algorithm is not affected by the cardinality of $I$ upon which the estimation of $z$ is carried on, providing remarkably good results for all the chosen values. In particular, our method is overall very accurate, being able to correctly recover the $80\%$ of 1s and the $73\%$ of 0s, a result to be compared with the performance of a perfect classifier, for which $\langle TPR\rangle=\langle SPC\rangle=1$, and with that of a random classifier, for which $\langle TPR\rangle=1-\langle SPC\rangle=c$ ($c$ being the link density of the whole network). The high accuracy of our reconstruction method is also witnessed by the low rate of false positives of our algorithm, due to the accurate estimation of the actual link density. As discussed in \cite{squartini2016stock}, overestimating the link density would have increased the expected $TPR$ (a method predicting a complete network is characterized by $\langle TPR\rangle=1$), at the price of increasing the rate of false positives as well, thus decreasing the predictive power of the method itself.

Our method performs well also in reproducing the weighted structure of the WTW: upon adding the correction term up to the third iteration of the IPF algorithm, the largest expected in-strength (reading $\langle s^{in}_{i,\:corr}\rangle=\sum_{j(\neq i)}\left(\frac{s^{out}_js^{in}_i}{W}+w_{ji}^{(3)}\right),\:\forall\:i$) accounts for the 95\% of the observed value. On the other hand, the non-corrected value $\langle s^{in}_{i}\rangle=\sum_{j(\neq i)}\left(\frac{s^{out}_js^{in}_i}{W}\right)$ accounts for the 82\% only. Better results are obtained for the out-strength sequence: the corrected value for the node characterized by the maximum out-strength amounts at the 99\% of the corresponding observed value (the non-corrected value accounts for the 88\%).

Overall, we obtain a value $\theta_{\text{WTW}}\simeq 0.712$ for all the considered cardinalities $n_I$, indicating a satisfactorily high level of similarity between our weights prediction and their observed values.

\subsection*{e-MID interbank network}

The second network we have tested our method upon is the electronic Market for Interbank Deposits (e-MID), i.e. the network whose nodes are banks and whose generic link $i\rightarrow j$ represents the loan granted from $i$ to $j$. We remand the reader to \cite{Iori2006} for more details on the dataset.

Table \ref{tab2} summarizes the results of our analysis on e-MID in the year 1999 only (again, similar results hold for the other years in our data set - see Appendix). As for the WTW, the performance of our algorithm is not affected by $n_I$ providing again very good results for the whole range of values of the subsets cardinality. In particular, our method is again very accurate, being able to correctly recover the $\simeq64\%$ of 1s and the $\simeq86\%$ of 0s. Even if the predictive power of our method is lower than for the WTW case, the accuracy values are comparable, amounting at $\simeq80\%$.

Our method performs also very well in reproducing the e-MID weighted structure: the correction term coming from the IPF algorithm and calculated for the maximum $\langle s^{out}_{i,\:corr}\rangle=\sum_{j(\neq i)}\left(\frac{s^{out}_is^{in}_j}{W}+w_{ij}^{(3)}\right),\:\forall\:i$ accounts for the 99\% of the observed value. On the other hand, the usual value $\langle s^{out}_{i}\rangle=\sum_{j(\neq i)}\left(\frac{s^{out}_is^{in}_j}{W}\right)$ accounts for the 88\% only. A comparable result is obtained for the in-strength sequence: the corrected value for the node characterized by the maximum in-strength still amounts at the 99\% of the corresponding observed value (the non-corrected value accounts for the 96\%).

The value $\theta_{\text{e-MID}}\simeq 0.82$ indicates that, on average, a very high level of similarity between observed and predicted weights is again obtained, confirming the degree-corrected gravity model as a good predictor of the links weights.

\begin{table*}[t!]
\[\begin{array}{l|c|c|c|c|c}
\hline
\text{Link density} & n=5\:[\text{CI}\:95\%] & n=10\:[\text{CI}\:95\%] & n=20\:[\text{CI}\:95\%] & n=50\:[\text{CI}\:95\%] & n=100\:[\text{CI}\:95\%] \\
\hline
\text{WTW 2000 (true: 0.578)} & 0.586\:[0.560;0.611] & 0.559\:[0.544;0.574] & 0.583\:[0.574;0.592] & 0.578\:[0.573;0.583] & 0.577\:[0.574;0.580] \\
\hline
\text{e-MID 1999 (true: 0.274)} & 0.292\:[0.271;0.313] & 0.278\:[0.267;0.289] & 0.276\:[0.268;0.283] & 0.276\:[0.272;0.280] & 0.275\:[0.273;0.278] \\
\hline
\end{array}\]
\caption{Link density estimation for different cardinalities $n$ of the random sampled subset $I$. Results are based on 1000 samples and are shown together with the $95\%$ confidence intervals. The considered cardinalities $n=5, 10, 20, 50, 100$ correspond to percentages ranging from $\simeq2\%$ to $\simeq50\%$ of the total number of nodes. The true link densities calculated on the entire networks are shown in brackets for reference.}\label{tab1}
\end{table*}

\subsection*{Random-nodes sampling scheme}

The sampling-based reconstruction algorithm we have proposed in the present paper rests upon the homogeneity assumption, according to which any subset of nodes picked at random provides a representative value of the density of the whole network. Table \ref{tab1} collects the estimations of the link density, averaged over all sampled subsets of a given cardinality: remarkably, the obtained values are accurate even for low cardinalities. In order to assess the magnitude of fluctuations, we have also explicitly computed the empirical probability distributions of the link density estimates, obtained by random sampling our nodes subsets. These distributions are shown in fig. \ref{fig1} (right panels). Naturally, the smaller the cardinality of the considered nodes subsets, the more spaced the values of the observable link density and the less smooth the corresponding probability distribution. These findings suggests that our homogeneity assumption is indeed verified, provided that nodes are sampled according to the random selection scheme \cite{Barrat2015}.

As a comparison, we have also sampled nodes sequentially, i.e. by, first, ordering nodes according to their total strength $s^{tot}_{i}=s^{out}_i+s^{in}_i$ and, then, considering bunches of $n$ subsequent nodes (again, for each value of $n$). For each subset of nodes we have calculated the corresponding internal link density and plotted it versus the total internal strength of nodes, i.e. $s^{tot}_{I}=\sum_{i\in I}\left(s^{out}_i+s^{in}_i\right)$. As shown in fig. \ref{fig1} (left panels), such a procedure provides insights on the structural organization of both WTW and e-MID: nodes characterized by large values of the total strength tend to form densely-connected groups whereas nodes characterized by small values of the total strength tend to form loosely-connected groups. Such an evidence confirms the presence of a core-periphery structure, with nodes having a smaller total strength establishing connections with nodes having a large total strength which, in turn, tend to connect preferentially with each other (as a sort of ``rich-club'') \cite{Fagiolo2010,DeMasi2006}. Our analysis suggests that a sampling-based reconstruction procedure must rest upon a ``balanced'' sampling of the nodes, biased neither  towards the ``core'' portion of nodes (which would lead to severely overestimate the overall network density), nor towards the ``periphery'' portion of nodes (which would lead to severely underestimate the overall network density). Interestingly, in a recent paper comparing several network sampling techniques was found that the least biased sampling scheme for estimating a given network density is precisely the random-nodes one \cite{Bagus2016}.

\section*{Conclusions}

The present contribution proposes a recipe to reconstruct a network from a very limited amount of information. In particular, we address the problem of inferring the binary and the weighted structure of a given network from the knowledge of the nodes strengths and the link density of only a subset of nodes. As we have shown in the paper, the best sampling scheme is the random-nodes selection scheme which ensures that an accurate estimation of the whole network density can indeed be achieved. On the contrary, selecting nodes on the basis of more informative structural properties (as the degree, or the strength) could bias the estimation of the connectance towards unrealistically too large, or too small, values. The role played by the available piece of topological information is fundamental not only to achieve an accurate reconstruction of the purely binary structure but also of the weighted structure, as evident upon inspecting table \ref{tab2}.

The aforementioned results have been obtained by estimating the link density of the whole network upon considering only nodes subsets: in other words, we have verified that different random subsets (even with different cardinality) are characterized by very similar densities, in turn implying that the whole network density can be estimated (with a high degree of accuracy) by considering a subset randomly drawn from the whole set of nodes. However, the proposed algorithm can be also used to reconstruct networks with a modular structure, upon tuning the link densities of the different modules via eq.(\ref{eq4}): examples are provided by interbank networks structured into jurisdictions, the latter playing the role of the subsets to be reconstructed.

\section*{Appendix 1}

A pseudo-code summarizing the main steps of the reconstruction algorithm presented in the paper follows.

\begin{algorithm}[H]
\DontPrintSemicolon
\SetAlgoNoLine
\SetKwInOut{Input}{Input}
\Input{in- and out-strengths $\{s_i^{in}\}_{i=1}^N$, $\{s_i^{out}\}_{i=1}^N$ and link density of a subset $I$, $c_I=\frac{L_I}{n_I(n_I-1)}$.}
\BlankLine
\Begin{
define $p_{ij}=\frac{zs_i^{out}s_j^{in}}{1+zs_i^{out}s_j^{in}}$;\;
solve the equation $c_I=\langle c_I\rangle$ in order to determine $z$:
\begin{equation*}
c_I=\frac{1}{n_I(n_I-1)}\sum_{i\in I}\sum_{j(\neq i)\in I}\frac{zs_i^{out}s_j^{in}}{1+zs_i^{out}s_j^{in}};\;
\end{equation*}
\For{$m=1\dots M$}{
\Indp\For{$i<j$}{
calculate the correction to the gravity-like estimation
\begin{equation*}
w_{ij}^{(3)}=\frac{s^{out}_is^{in}_i}{W}\left[\frac{s^{out}_js^{in}_j}{\sum_{l(\neq j)}s^{out}_ls^{in}_l}\right]\left[\frac{1}{\sum_{k(\neq i)}\frac{s^{out}_ks^{in}_k}{\sum_{m(\neq k)}s^{out}_ms^{in}_m}}\right];\;
\end{equation*}
connect $i$ and $j$ with a weight drawn from the following Bernoulli distribution\;
\begin{equation*}
w_{ij} = \left\{ \begin{array}{cl}
0, & \textrm{$1-p_{ij}$},\\
\left(\frac{s_i^{out}s_j^{in}}{W}+w_{ij}^{(3)}\right)\frac{1}{p_{ij}}, & \textrm{$p_{ij}$}.
\end{array} \right.
\end{equation*}
}}
verify the goodness of the achieved reconstruction by calculating the ensemble average of indicators like $TPR$, $SPC$, $PPV$, $ACC$ and $\theta$;
}
\BlankLine
\SetKwInOut{Output}{Output}
\Output{ensemble of $M$ reconstructed directed, weighted networks.}
\caption{Network reconstruction via density sampling}
\end{algorithm}

\section*{Appendix 2}

Additional years have been analysed for both the WTW and e-MID (see tables \ref{tab3} and \ref{tab4}).

\begin{table*}[t!]
\[\begin{array}{l|c|c|c|c}
\hline
\hline
\text{\bf WTW} & n=5\:(\text{CI}\:95\%) & n=10\:(\text{CI}\:95\%) & n=20\:(\text{CI}\:95\%) & n=50\:(\text{CI}\:95\%) \\
\hline
\hline
\text{1950 - Link density (true: 0.402)} & 0.401\:[0.375;0.426] & 0.402\:[0.387;0.416] & 0.401\:[0.393;0.409] & 0.400\:[0.396;0.403] \\
\hline
\text{1950 - Accuracy} & 0.736\:[0.731;0.741] & 0.747\:[0.746;0.749] & 0.751 & 0.752 \\
\hline
\text{1950 - Cosine similarity} & 0.460\:[0.458;0.462] & 0.463 & 0.463 & 0.463 \\
\hline
\hline
\text{1960 - Link density (true: 0.383)} & 0.329\:[0.305;0.352] & 0.343\:[0.330;0.357] & 0.346\:[0.338;0.355] & 0.348\:[0.344;0.353] \\
\hline
\text{1960 - Accuracy} & 0.737\:[0.734;0.741] & 0.746 & 0.749\:[0.748;0.750] & 0.751 \\
\hline
\text{1960 - Cosine similarity} & 0.586 & 0.591 & 0.591 & 0.591 \\
\hline
\hline
\text{1970 - Link density (true: 0.460)} & 0.464\:[0.436;0.492] & 0.478\:[0.462;0.496] & 0.461\:[0.451;0.471] & 0.464\:[0.458;0.469] \\
\hline
\text{1970 - Accuracy} & 0.695\:[0.691;0.699] & 0.704\:[0.702;0.706] & 0.709 & 0.709 \\
\hline
\text{1970 - Cosine similarity} & 0.669 & 0.669 & 0.669 & 0.669 \\
\hline
\hline
\text{1980 - Link density (true: 0.468)} & 0.484\:[0.458;0.510] & 0.470\:[0.455;0.485] & 0.471\:[0.461;0.481] & 0.463\:[0.458;0.469] \\
\hline
\text{1980 - Accuracy} & 0.719\:[0.715;0.723] & 0.731\:[0.730;0.733] & 0.734\:[0.733;0.735] & 0.736 \\
\hline
\text{1980 - Cosine similarity} & 0.732 & 0.732 & 0.732 & 0.732 \\
\hline
\hline
\text{1990 - Link density (true: 0.505)} & 0.495\:[0.467;0.522] & 0.516\:[0.500;0.532] & 0.506\:[0.497;0.515] & 0.507\:[0.503;0.512] \\
\hline
\text{1990 - Accuracy} & 0.731\:[0.726;0.736] & 0.743\:[0.741;0.745] & 0.748 & 0.749 \\
\hline
\text{1990 - Cosine similarity} & 0.751 & 0.751 & 0.751 & 0.751 \\
\hline
\end{array}\]
\caption{Statistical indicators used to evaluate the performance of our sampled-based reconstruction method, for different cardinalities $n$ of the known subset $I$. Results are shown together with the $95\%$ confidence intervals (not shown whenever their difference affects the significant digits beyond the third one). The considered cardinalities $n=5, 10, 20, 50$ correspond to percentages ranging from $\simeq2\%$ to $\simeq25\%$ of the total number of nodes. The true link densities calculated on the entire networks for the various periods are shown in brackets for reference.}\label{tab3}
\end{table*}

\begin{table*}[t!]
\[\begin{array}{l|c|c|c|c}
\hline
\hline
\text{\bf e-MID} & n=5\:(\text{CI}\:95\%) & n=10\:(\text{CI}\:95\%) & n=20\:(\text{CI}\:95\%) & n=50\:(\text{CI}\:95\%) \\
\hline
\hline
\text{2000 - Link density (true: 0.278)} & 0.293\:[0.269;0.317] & 0.279\:[0.263;0.295] & 0.273\:[0.264;0.281] & 0.280\:[0.273;0.283] \\
\hline
\text{2000 - Accuracy} & 0.763\:[0.759;0.768] & 0.772\:[0.769;0.775] & 0.778\:[0.777;0.779] & 0.778\:[0.777;0.779] \\
\hline
\text{2000 - Cosine similarity} & 0.573\:[0.566;0.580] & 0.578\:[0.576;0.582] & 0.582 & 0.582 \\
\hline
\hline
\text{2001 - Link density (true: 0.263)} & 0.279\:[0.256;0.303] & 0.264\:[0.249;0.278] & 0.257\:[0.246;0.267] & 0.266\:[0.261;0.272] \\
\hline
\text{2001 - Accuracy} & 0.763\:[0.757;0.770] & 0.774\:[0.772;0.776] & 0.777\:[0.775;0.779] & 0.778\:[0.777;0.779] \\
\hline
\text{2001 - Cosine similarity} & 0.560\:[0.554;0.566] & 0.566\:[0.563;0.569] & 0.569 & 0.570 \\
\hline
\hline
\text{2002 - Link density (true: 0.233)} & 0.253\:[0.230;0.276] & 0.237\:[0.221;0.252] & 0.235\:[0.225;0.246] & 0.233\:[0.228;0.239] \\
\hline
\text{2002 - Accuracy} & 0.759\:[0.752;0.766] & 0.767\:[0.763;0.771] & 0.770\:[0.767;0.772] & 0.772\:[0.770;0.773] \\
\hline
\text{2002 - Cosine similarity} & 0.684\:[0.675;0.694] & 0.670\:[0.682;0.697] & 0.699\:[0.697;0.701] & 0.701\:[0.700;0.702] \\
\hline
\hline
\text{2003 - Link density (true: 0.214)} & 0.248\:[0.223;0.273] & 0.225\:[0.208;0.243] & 0.217\:[0.205;0.228] & 0.213\:[0.208;0.219] \\
\hline
\text{2003 - Accuracy} & 0.746\:[0.737;0.756] & 0.758\:[0.752;0.763] & 0.763\:[0.759;0.766] & 0.766\:[0.764;0.767] \\
\hline
\text{2003 - Cosine similarity} & 0.4614\:[0.453;0.470] & 0.462\:[0.454;0.470] & 0.472\:[0.469;0.475] & 0.476\:[0.475;0.477] \\
\hline
\hline
\text{2004 - Link density (true: 0.190)} & 0.210\:[0.185;0.235] & 0.183\:[0.168;0.199] & 0.194\:[0.182;0.205] & 0.187\:[0.181;0.192] \\
\hline
\text{2004 - Accuracy} & 0.772\:[0.762;0.783] & 0.785\:[0.780;0.790] & 0.784\:[0.780;0.788] & 0.788\:[0.786;0.790] \\
\hline
\text{2004 - Cosine similarity} & 0.481\:[0.470;0.492] & 0.482\:[0.474;0.491] & 0.497\:[0.493;0.502] & 0.503\:[0.501;0.504] \\
\hline
\hline
\text{2005 - Link density (true: 0.201)} & 0.232\:[0.205;0.258] & 0.210\:[0.190;0.222] & 0.210\:[0.200;0.221] & 0.208\:[0.203;0.214] \\
\hline
\text{2005 - Accuracy} & 0.751\:[0.740;0.762] & 0.767\:[0.760;0.773] & 0.767\:[0.763;0.771] & 0.769\:[0.767;0.771] \\
\hline
\text{2005 - Cosine similarity} & 0.461\:[0.448;0.474] & 0.476\:[0.470;0.483] & 0.486\:[0.483;0.490] & 0.491\:[0.490;0.492] \\
\hline
\end{array}\]
\caption{Statistical indicators used to evaluate the performance of our sampled-based reconstruction method, for different cardinalities $n$ of the known subset $I$. Results are shown together with the $95\%$ confidence intervals (not shown whenever their difference affects the significant digits beyond the third one). The considered cardinalities $n=5, 10, 20, 50, 100$ correspond to percentages ranging from $\simeq2\%$ to $\simeq25\%$ of the total number of nodes. The true link densities calculated on the entire networks for the various periods are shown in brackets for reference.}\label{tab4}
\end{table*}

\section*{Competing Interests}
The authors declare that they have no competing interests.

\section*{Acknowledgments}
This work was supported by the EU projects CoeGSS (grant num. 676547), Multiplex (grant num. 317532), Shakermaker (grant num. 687941), SoBigData (grant num. 654024) and the FET projects SIMPOL (grant num. 610704), DOLFINS (grant num. 640772). DG acknowledges support from the Econophysics foundation (Stichting Econophysics, Leiden, the Netherlands).

\section*{Authors' Contributions}
TS, GC, AG and DG participated in the design of the analysis. TS and GC performed the statistical analysis. All authors wrote, read and approved the final manuscript.

\bibliography{bibfile_main}

\end{document}